# SSL-QALAS: Self-Supervised Learning for Rapid Multiparameter Estimation in Quantitative MRI Using 3D-QALAS

Yohan Jun[1,2], Jaejin Cho[1,2], Xiaoqing Wang[1,2], Michael Gee[2,3], P. Ellen Grant[2,4], Berkin Bilgic[1,2,5,*], and Borjan Gagoski[2,4,*]

[1] Athinoula A. Martinos Center for Biomedical Imaging, Charlestown, MA, United States

[2] Department of Radiology, Harvard Medical School, Boston, MA, United States

[3] Department of Radiology, Massachusetts General Hospital, Boston, MA, United States

[4] Fetal-Neonatal Neuroimaging & Developmental Science Center, Boston Children's Hospital, Boston, MA, United States

[5] Harvard/MIT Health Sciences and Technology, Cambridge, MA, United States

* Equal contribution as last authors

Corresponding Author

Yohan Jun

Athinoula A. Martinos Center for Biomedical Imaging, Charlestown, MA, United States

Email: yjun@mgh.harvard.edu

Grant Support: This work was supported by research grants NIH R01 EB032378, R01 EB028797, R03 EB031175, U01 EB025162, P41 EB030006, U01 EB026996, U01 DA055353 and the NVIDIA Corporation for computing support.

Running Title: Self-Supervised Learning for Rapid Multiparametric Estimation in Quantitative MRI Using 3D-QALAS

**Abstract**

**Purpose:** To develop and evaluate a method for rapid estimation of multiparametric $T_1$, $T_2$, proton density (PD), and inversion efficiency (IE) maps from 3D-quantification using an interleaved Look-Locker acquisition sequence with $T_2$ preparation pulse (3D-QALAS) measurements using self-supervised learning (SSL) without the need for an external dictionary.

**Methods:** A SSL-based QALAS mapping method (SSL-QALAS) was developed for rapid and dictionary-free estimation of multiparametric maps from 3D-QALAS measurements. The accuracy of the reconstructed quantitative maps using dictionary matching and SSL-QALAS was evaluated by comparing the estimated $T_1$ and $T_2$ values with those obtained from the reference methods on an ISMRM/NIST phantom. The SSL-QALAS and the dictionary matching methods were also compared *in vivo*, and generalizability was evaluated by comparing the scan-specific, pre-trained, and transfer learning models.

**Results:** Phantom experiments showed that both the dictionary matching and SSL-QALAS methods produced $T_1$ and $T_2$ estimates that had a strong linear agreement with the reference values in the ISMRM/NIST phantom. Further, SSL-QALAS showed similar performance with dictionary matching in reconstructing the $T_1$, $T_2$, PD, and IE maps on *in vivo* data. Rapid reconstruction of multiparametric maps was enabled by inferring the data using a pre-trained SSL-QALAS model within 10 s. Fast scan-specific tuning was also demonstrated by fine-tuning the pre-trained model with the target subject's data within 15 min.

**Conclusion:** The proposed SSL-QALAS method enabled rapid reconstruction of multiparametric maps from 3D-QALAS measurements without an external dictionary or labeled ground-truth training data.

**INTRODUCTION**

Quantitative MRI (qMRI) provides quantitative tissue information, including $T_1$, $T_2$, $T_2^*$ relaxation rates, and proton density (PD) estimates, which can be used for tissue analysis to help improve clinical diagnosis (1–11) in diseases such as brain tumors and multiple sclerosis. In addition, such information also finds applications in neuroscientific imaging, including aging research (9,12–18). Quantitative information can be obtained using sequences including DESPOT1, DESPOT2, MP2RAGE, and MPnRAGE, which acquire a single quantitative magnetic resonance (MR) parameter map from the measurements (19–21). Recently, various MRI methods that have been developed are capable of acquiring multiple quantitative MR parameter maps at high resolution within a reasonable timeframe, including magnetic resonance fingerprinting (MRF) (22), echo planar time-resolved imaging (EPTI) (23), and MR multitasking (24). 3D MRF, which obtains $T_1$, $T_2$, and PD maps using a tailored acquisition scheme with a spiral k-space trajectories, showed high repeatability and reproducibility of multiparametric maps in whole-brain imaging at 1.125 mm isotropic resolution within 10 min (25,26). 3D EPTI combines an inversion-recovery gradient-echo (IR-GE) and variable flip angle (VFA) gradient and spin echo (GRASE) acquisition schemes and integrates them with controlled aliasing in parallel imaging (CAIPI), enabling whole-brain $T_1$, $T_2$, $T_2^*$, PD, and $B_1^+$ maps at 1 mm isotropic resolution within 3 min (23). MR multitasking using $T_2$-prepared inversion-recovery and multi-echo gradient-echo (mGRE) readouts enabled whole-brain 3D acquisition of $T_1$, $T_2$, and $T_2^*$ maps at $0.7 \times 1.4 \times 2$ mm$^3$ resolution within 9.1 min (27).

Recently, 3D-quantification using an interleaved Look-Locker acquisition sequence with $T_2$ preparation pulse (3D-QALAS) has been developed for acquiring high-resolution $T_1$, $T_2$, and PD maps from five measurements within each repetition time (TR) (28–30). The original 3D QALAS was proposed for acquiring rapid $T_1$ and $T_2$ maps in cardiac imaging (28,31,32). It has been applied to neuroimaging (29,30,33), enabling whole-brain $T_1$, $T_2$, and PD maps at 1 mm isotropic resolution within 11 min of acquisition time. Recent studies evaluating 3D-QALAS showed reliable scan-rescan repeatability for measuring cortical thickness and subcortical brain volumes, as well as high repeatability of the estimated $T_1$, $T_2$, and PD values (29,30). These maps can be further utilized to synthesize multiple contrast weighted volumes used in clinical settings, such as $T_1$-weighted ($T_1$w), $T_2$-weighted ($T_2$w), fluid-attenuated inversion-recovery (FLAIR), phase-sensitive inversion recovery (PSIR), and double inversion recovery (DIR), which also can help evaluate multiple sclerosis lesions (14).

There have been efforts to reduce the scan time of QALAS using compressed sensing (CS), which enabled an additional 2-fold acceleration of "conventional" QALAS (acquired at $R = 2$ using parallel imaging), while maintaining equivalent quantitative values of obtained maps and similar image quality of synthesized images (33,34). Recent efforts have also employed wave-controlled aliasing in parallel imaging (Wave-CAIPI) (35,36) to accelerate the QALAS scan by 6-fold using corkscrew k-space trajectories and generalized parallel imaging reconstruction. This acquisition was further combined with model-based deep learning reconstruction to push the acceleration to 12-fold (37).

While those methods could reduce the image acquisition time by reconstructing the multi-contrast images of QALAS from undersampled k-space data, an additional fitting process is still needed to generate quantitative maps, which requires additional computation time. Bloch-simulation-based dictionary matching is one way to generate the quantitative maps; however, it requires an external dictionary that needs to be pre-calculated, and it requires a long computation time to perform the voxel-by-voxel fitting, especially for high-resolution images, which might hinder online reconstruction. A similar problem is faced in MRF, where advanced computational algorithms for dictionary generation or signal matching were presented, such as using GPU acceleration (38), fast group matching (39), and dimensionality reduction of dictionaries using a low-rank approximation (40) and singular value decomposition (SVD) (41).

Along with the recent rapid growth of deep-learning-based algorithms, many studies have exploited those methods for MRI applications (42–49). For example, 3DMRF-DL, which combined parallel imaging and deep learning, demonstrated that high-resolution whole-brain $T_1$ and $T_2$ maps could be acquired at 1 mm isotropic resolution within 7 min (48). Further, MANTIS, a model-augmented neural network with incoherent k-space sampling, showed rapid $T_2$ mapping using a multi-echo spin-echo (ME-SE) sequence which was accelerated up to 8-fold (49). While the deep-learning-based models, which were trained in a supervised way, outperformed conventional algorithms in numerous applications, there are various cases where ground-truth or label images are challenging to be defined or acquired. In the case of quantitative mapping, the acquisition of *in vivo* gold standard quantitative maps at high resolution may be infeasible using conventional scans such as inversion-recovery spin-echo (IR-SE) or fast-spin-echo (IR-FSE) for $T_1$, and single-echo spin-echo (SE-SE) or fast-spin-echo (SE-FSE) for $T_2$ maps, as these often suffer from prohibitively long scan times.

On the other hand, self-supervised learning (SSL) does not require external training data for model training and can be used in denoising, reconstruction, and quantitative mapping (50–55). Several SSL-based qMRI studies have been proposed for $T_1$ mapping using variable flip angle (VFA) imaging with a spoiled gradient-echo (SPGR) sequence (53), $T_1\rho$ mapping with various time of spin-lock (TSL) pulses (54), $T_2$ mapping using a ME-SE sequence (53), and $R_2^*$ mapping using an mGRE sequence (55). While these studies generated a single quantitative map, there have been only a few studies for multiparametric mapping (56,57).

In this study, we propose to estimate multiple quantitative maps, including $T_1$, $T_2$, PD, and inversion efficiency (IE) maps, by employing the SSL methods to 3D-QALAS measurements (i.e., **SSL-QALAS**), enabling rapid and dictionary-free multiparametric fitting, which is trained in a scan-specific way. We validated our method using an International Society for Magnetic Resonance in Medicine and National Institute of Standards and Technology (ISMRM/NIST) system phantom, where our proposed SSL-QALAS model obtained a high correlation with the reference methods (IR-FSE and SE-FSE) in $T_1$ and $T_2$ maps. We also demonstrated that SSL-QALAS showed similar performance with the dictionary matching method in reconstructing the $T_1$, $T_2$, PD, and IE maps using *in vivo* data. The rapid reconstruction of the multiparametric maps was enabled by inferring the data using a pre-trained SSL-QALAS model, which provided up to 360-fold computational speedup compared to dictionary matching (within 10 s). Fast scan-specific improvements were also demonstrated by fine-tuning the pre-trained model with the target subject's data (within 15 min). All the source codes can be found here: https://github.com/yohan-jun/SSL-QALAS

## METHODS

### Self-Supervised Learning for Multiparametric Mapping

The overall flowchart of the proposed SSL-QALAS method for multiparametric quantitative mapping is presented in Fig. 1a. SSL-QALAS is based on convolutional neural network (CNN) architecture. The five acquired QALAS contrast images (**y**) along with a separately acquired $B_1^+$ map are fed into the CNN model ($\mathcal{D}(\cdot\,;\boldsymbol{\theta})$) as the input, where $\boldsymbol{\theta}$ denotes the trainable parameters of the model $\mathcal{D}$. The model then estimates $T_1$, $T_2$, PD, and IE maps as the output ($\mathcal{D}(\mathbf{y}, B_1^+; \boldsymbol{\theta})$). The loss function calculates the $l_2$ loss ($\mathcal{L}$) between the acquired images and synthetic images, which are generated by feeding the output maps into the forward signal model $S$:

$$\min_{\boldsymbol{\theta}} \mathcal{L}\left(\mathbf{y}, S(\mathcal{D}(\mathbf{y}, B_1^+; \boldsymbol{\theta}))\right) = \min_{\boldsymbol{\theta}} \left\|\mathbf{y} - S(\mathcal{D}(\mathbf{y}, B_1^+; \boldsymbol{\theta}))\right\|_2^2, \qquad [1]$$

where $S$ denotes the QALAS signal model, which is dependent on $T_1$, $T_2$, PD, IE, and $B_1^+$ maps. After the training of the model $\mathcal{D}$ is finished, $T_1$, $T_2$, PD, and IE maps can be inferred using the optimized parameters $\boldsymbol{\theta}$.

**Implementation Details**

*SSL-QALAS*

The architecture of the SSL-QALAS network is presented in Supporting Information Figure S1. The five acquired QALAS contrast images were concatenated along the channel dimension before feeding them into the CNN model. The proposed architecture consists of 5 CNN blocks where each layer has a 1×1 convolutional layer with 64 feature maps, instance normalization layer (58), and leaky rectified linear unit (ReLU) activation function. The last block has a sigmoid activation function with constant multiplication for generating 4 different outputs, i.e., $T_1$, $T_2$, PD, and IE maps, which helps the output to be in user-specified physical ranges. The model was trained with Adam optimizer (59) with $\beta_1 = 0.9$ and $\beta_2 = 0.999$ for 500 epochs with a learning rate of 0.001. For implementing the QALAS signal model $S$, we followed the forward equations of the original QALAS paper (28). The deep learning model and the QALAS signal model were implemented using the PyTorch library (60). The training of the model from scratch using a single subject's multi-slice data took about 1–1.5 h, depending on the matrix size of the image, while fine-tuning the model took about 10–15 min using a single NVIDIA RTX A6000 GPU. The inference using the pre-trained model took about 40–50 ms for each slice (less than 10 s for the whole volume) using the same GPU. To accelerate the model training, the validation step was conducted every 10 epochs, using batch-size 4.

*Dictionary Matching*

The dictionary was generated based on QALAS signal model (28) with the following $T_1$, $T_2$, and IE ranges: $T_1$ = [5–5000 ms], $T_2$ = [1–2500 ms], and IE = [0.5–1.0]. A small step size was used for short $T_1$ and $T_2$ values while it was increased gradually for long $T_1$ and $T_2$ values. Specifically, for $T_1$ values, a 5 ms step size was used for 5–3000 ms and 100 ms for 3000–5000 ms. For $T_2$ values, a 2 ms step size was used for 1–350 ms, 20 ms for 350–1000 ms, and 200 ms for 1000–2500 ms. In addition, a 0.02 step size was used for 0.5–1.0 IE values. The sequence diagram of QALAS is presented in Fig. 1b. The dictionary generation and matching of the whole volume took ~30 min and 1 h, respectively, using 16 CPU cores and MATLAB Parallel Computing Toolbox, which allowed parallel processing of multiple slices.

**Image Acquisition**

*Phantom Experiments*

To validate the $T_1$ and $T_2$ accuracy of the 3D-QALAS sequence, phantom experiments were conducted using an ISMRM/NIST system phantom (Serial Number 0127) on a 3T MAGNETOM Prisma scanner (Siemens Healthineers, Erlangen, Germany) with a 32ch head receive array. For $B_1^+$ inhomogeneity correction, $B_1^+$ maps were acquired using a separate turbo-fast low-angle shot (turbo-FLASH) sequence (61). The reference $T_1$ and $T_2$ maps were acquired using multiple IR-FSE and SE-FSE scans, respectively. Detailed imaging parameters of the sequences used for the phantom experiments can be found in Supporting Information Tables S1 and S2.

*In vivo Experiments*

The *in vivo* experiments were conducted with the approval of the Institutional Review Board. *In vivo* data were acquired from a healthy volunteer (subject #1; male; age, 33) using a 3D-QALAS sequence using the same 3T scanner and head receive array that was used for the phantom experiments. To evaluate repeatability, four 3D-QALAS runs were acquired on the same volunteer: 'initial scan', 're-scan' (without any changes), 're-landmarked scan' (altering the landmark position), and 're-positioned scan' (the volunteer was asked to get up from the patient table and was re-brought back in). To correct $B_1^+$ inhomogeneity, $B_1^+$ maps were separately acquired using a turbo-FLASH sequence (61) for each 3D-QALAS acquisition. Detailed imaging parameters of the sequences used in the *in vivo* repeatability experiments can be found in Supporting Information Table S1 ('*In vivo* Experiment #1').

To further validate the generalization of the proposed SSL-QALAS model, additional *in vivo* data were acquired from ten healthy volunteers (subjects #2–11; 7 males and 3 females; age range, 22–50 years) using the same 3D-QALAS and turbo-FLASH sequences. In all experiments, the $B_1^+$ maps were interpolated to have the same matrix size as the 3D-QALAS images, and maximum and minimum values were thresholded to have a range of 0.65 to 1.35. Detailed imaging parameters of the sequences used in the *in vivo* generalizability experiments can be found in Supporting Information Table S1 ('*In vivo* Experiment #2').

**Model Comparisons**

To validate the accuracy of the $T_1$ and $T_2$ maps estimated by the dictionary matching and our proposed SSL-QALAS methods from 3D-QALAS measurements, the reconstructed $T_1$ and $T_2$ maps were compared with the reference maps using a linear regression method. From the $T_2$ plate of the ISMRM/NIST system phantom, spheres with $T_1$ and $T_2$ values within the physiological values of the adult brain tissues (eight spheres with $T_1$ values between 600–3200 ms and six spheres with $T_2$ values between 40–260 ms), were analyzed, by measuring the mean values of the circular region of interests (ROI) drawn inside the spheres using the ITK-SNAP software (https://www.itksnap.org/) (62).

The *in vivo* analysis compared the estimated $T_1$, $T_2$, PD, and IE maps generated using the dictionary matching and SSL-QALAS methods. The repeatability of the SSL-QALAS method was validated by comparing the maps reconstructed using a scan-specific trained model and a pre-trained model with the initial data. In the scan-specific model, the training and inference data are identical. The pre-trained model was trained with the initial data and rapidly inferred using data from the other acquisitions, including re-scan, re-landmark, and re-position.

Furthermore, the generalizability and reproducibility of the SSL-QALAS method were validated by comparing the maps reconstructed using a subject-specific (subject #3–11) trained model, a pre-trained model with the subject #2's data, and a transfer learning model where the model was initially trained with the subject #2's data and fine-tuned with the target subject #3–11's data. The reconstructed images were evaluated by the root mean square error (RMSE) metric.

**RESULTS**

**Phantom Evaluation**

*Accuracy Evaluation of 3D-QALAS using Dictionary Matching and SSL-QALAS*

The quantitative analyses of 3D-QALAS using ISMRM/NIST system phantom are presented in Fig. 2. A strong linear agreement of $T_1$ and $T_2$ values within physiologically reasonable ranges for adult brain's tissues was observed between 3D-QALAS and reference methods (IR-FSE and SE-FSE) where the dictionary matching showed the coefficient of determination ($R^2$) = 0.9994 and 0.9997 and SSL-QALAS showed $R^2$ = 0.9998 and 0.9996 for $T_1$ and $T_2$ values, respectively. In addition, the dictionary matching showed the regression coefficient (i.e., slope of the fitted line) 1.0008 and 1.0496, and SSL-QALAS showed 1.0113 and 0.9595 for $T_1$ and $T_2$ values, respectively. Similar results were obtained when comparing the linear agreement between the dictionary matching and SSL-QALAS methods ($R^2$ = 0.9998 and 0.9999, and regression coefficient 0.9875 and 0.9926, for $T_1$ and $T_2$ values, respectively). The results for validating the repeatability of 3D-QALAS and generalizability of SSL-QALAS are presented in Supporting Information Figures S2 and S3.

### *In vivo* Evaluation

*Comparison Between Dictionary Matching and SSL-QALAS*

The reconstructed $T_1$, $T_2$, PD, and IE maps of *in vivo* data using the dictionary matching and the scan-specific SSL-QALAS methods are presented in Fig. 3. As shown, the scan-specific SSL-QALAS method produces comparable quantitative maps with those obtained using dictionary matching. The RMSE values, calculated between the dictionary matching and the scan-specific SSL-QALAS methods in brain parenchyma regions where cerebrospinal fluid (CSF) regions are excluded, are 8.62, 8.75, 9.98, and 2.42% for $T_1$, $T_2$, PD, and IE, respectively.

*Generalizability of SSL-QALAS*

Fig. 4 shows $T_1$ and $T_2$ maps (of a representative axial slice) from 4 different SSL-QALAS acquisitions (initial scan, re-scan, re-landmark, and re-position) done on the same subject. The pre-trained model with the initial scan data shows comparable $T_1$ and $T_2$ maps with those of the scan-specific model except for some CSF regions, which are known to have very long $T_1$ and $T_2$ values compared to other tissues (see the difference images). The RMSE values, calculated between the scan-specific and the pre-trained models for each scan data, are 5.59, 6.08, and 4.68 % for $T_1$ and 11.68, 22.85, and 8.50 % for $T_2$. For further validation, SSL-QALAS models (i.e., scan-specific and pre-trained models) were compared with the dictionary matching method, which is presented in Supporting Information Figures S4. Moreover, the repeatability results of the SSL-QALAS method using 4 different acquisitions were presented in Supporting Information Figure S5.

In addition, the reconstructed $T_1$ and $T_2$ maps of 3 different SSL-QALAS models, including scan-specific, pre-trained, and transfer learning models, are presented in Fig. 5. Here, the results of a representative subject (subject #3) are presented, and quantitative results using 10 subjects are presented in Supporting Information Tables S3 and S4. The scan-specific model takes about 1.5 h to train the model from scratch. The pre-trained model, which only requires an inference process, takes about 10 s but shows some bias compared to the scan-specific model, as shown in the difference images. The transfer learning model, which requires both fine-tuning and inference processes, takes about 15 min for the reconstruction but shows reduced errors compared to the pre-trained model, as the RMSE values decreased from 6.28 and 18.85 % to 4.07 and 9.10 % for $T_1$ and $T_2$, respectively. Moreover, when the CSF regions are excluded from the calculations, the RMSE values decreased from 4.90 and 7.35 % to 2.88 and 4.47 % for $T_1$ and $T_2$, respectively. Similar results were observed in PD and IE maps, which are provided in Supporting Information Figure S6. For further validation, 3 different SSL-QALAS models (i.e., scan-specific, pre-trained, and transfer learning models) were compared with the dictionary matching method, which is presented in Supporting Information Figures S7 and S8.

Moreover, the reconstructed quantitative maps, including $T_1$, $T_2$, PD, and IE maps, showed reduced blurring by regularizing the maps using an additional total variation loss, which the dictionary matching method is infeasible to do, as shown in Fig. 6.

**DISCUSSION**

In this study, we proposed an SSL-QALAS method that estimates multiparametric quantitative $T_1$, $T_2$, PD, and IE maps from 3D-QALAS measurements, which trains a model without external labeled or ground-truth data and does not require an external dictionary for multiparametric fitting. While several SSL-based studies have been proposed for quantifying single parameter maps such as $T_1$ (53), $T_1\rho$ (54), $T_2$ (53), and $R_2^*$ (55), there have been only a few studies for multiparametric mapping using an SSL technique (56,57).

The main advantage of our proposed SSL-QALAS method is a rapid reconstruction of the multiparametric maps from QALAS measurements without an external dictionary while obtaining comparable quantitative results with the conventional dictionary matching method. The dictionary matching that we compared in the experiments requires 1.5 h, including dictionary generation, to reconstruct each subject's data. The computational cost of the proposed scan-specific SSL-QALAS model is on par with our implementation of the dictionary matching method, even when the scan-specific model is trained from scratch. Moreover, the proposed SSL-QALAS enables us to rapidly reconstruct the quantitative maps within 10 s using a pre-trained model, and also those can be improved by scan-specific fine-tuning the model within 15 min.

In our experiments using an ISMRM/NIST system phantom, the quantitative $T_1$ and $T_2$ values acquired from 3D-QALAS using both the dictionary matching ($R^2$ = 0.9994 and 0.9997) and SSL-QALAS ($R^2$ = 0.9998 and 0.9996) methods showed strong agreement with the reference values acquired from IR-FSE and SE-FSE. There was also a high correlation between the dictionary matching and SSL-QALAS in both $T_1$ and $T_2$ values ($R^2$ = 0.9998 and 0.9999), which were similarly shown in the reconstructed *in vivo* maps. This demonstrates that the proposed SSL-QALAS can estimate multiparametric maps from 3D-QALAS measurements with high accuracy. In addition, it also demonstrates that both methods had a strong agreement with the reference methods ($R^2$ > 0.999 and 0.95 < regression coefficient < 1.05) and showed comparable results. Since both methods demonstrated strong agreements with the reference methods and worked equally well, as shown in the phantom experiments, the dictionary matching is not necessarily more accurate than the SSL-QALAS method; thus, the dictionary matching does not have to be taken as ground truth necessarily.

The proposed SSL-QALAS method can be trained using the signal model based on the QALAS sequence timings and the five acquired sets of multi-contrast images, and it does not require an external pre-calculated dictionary. The model can be trained in a scan-specific way where the data for training and inference are identical. Since the model is trained generally to estimate the maps from 3D-QALAS measurements by solving the non-linear signal model using automatic differentiation with a neural network, the pre-trained model can be applied to other subject's data for rapid inference. While the pre-trained model showed some biases compared to the scan-specific model on our *in vivo* data, they could be reduced by fine-tuning the model using the target data, which took less time than training the model from scratch. Specifically, the RMSE values of scan-rescan were 5.59 and 11.68 % for $T_1$ and $T_2$, respectively, as shown in Fig. 4, whereas those of transfer learning model were 4.07 and 9.10 %, as shown in Fig. 5, which suggests that the error becomes comparable to or slightly lower than applying the pre-trained model directly on the rescan of the same subject.

In the SSL-QALAS model comparison, high RMSE values were obtained due to CSF regions, and they were much lower in the brain parenchyma when CSF regions were excluded. While both the dictionary matching and SSL-QALAS methods showed strong agreement with the reference values acquired from IR-FSE and SE-FSE within physiologically reasonable ranges for adult brain tissues ($T_1$: 600–3200 ms and $T_2$: 40–260 ms), the phantom experiments did not include very long $T_1$ and $T_2$ values such as CSF regions, which are known to have up to 4 s and 2 s for $T_1$ and $T_2$ values, respectively. Likewise, the phantom experiments did not validate for very long $T_1$ and $T_2$ values where those values are known to correspond to CSF regions; thus, the deviation between the phantom and *in vivo* experiments might occur, especially for those regions. In future works, further validation for the phantom experiments, including very long $T_1$ and $T_2$ values, would be needed.

Moreover, while our results demonstrated that spatial convolution (e.g., 3×3 convolutional layers) did not outperform voxel-wise convolution in terms of RMSE values in the generalizability experiments (in Supporting Information Figures S9 and S10), other architecture or loss function design might be needed to increase the signal-to-noise ratio of the reconstructed quantitative maps. For instance, by regularizing the quantitative maps, including $T_1$, $T_2$, PD, and IE maps, using an additional total variation loss, the reconstructed maps showed mitigated blurring (Fig. 6). Moreover, the proposed SSL-QALAS method can deal with different matrix sizes and resolutions of the data where the results were demonstrated in Supporting Information Figures S11 and S12.

The dictionary matching method estimated $T_1$, $T_2$, and IE values first, then estimated PD value by finding the optimal weighting value, which minimizes the difference between the simulated signals using the QALAS signal model with the estimated $T_1$, $T_2$, and IE values and the acquired QALAS signals. For $B_1^+$ correction, the dictionary matching was conducted by discretizing the $B_1^+$ maps (ranging from 0.65 to 1.35) into 100 bins to generate sub-dictionaries with each of the discretized $B_1^+$ values. In contrast, the SSL-QALAS model jointly estimated $T_1$, $T_2$, PD, and IE values at once from the acquired QALAS images using the acquired $B_1^+$ maps without the need for discretization. While the dictionary matching and SSL-QALAS methods employ different estimation approaches, the way of estimating $T_1$, $T_2$, PD, and IE values using the dictionary matching method is not necessarily more robust to $B_1^+$ effects compared to the SSL-QALAS method since the results using the SSL-QALAS method demonstrated comparable PD maps with the ones of the dictionary matching method, as shown in Fig. 3.

The accuracy of the $T_1$ estimation in sequences that use an inversion RF pulse can be affected by the incomplete inversion of longitudinal magnetization due to $T_2$ relaxation when using a lengthy adiabatic inversion pulse. This effect is quantified by IE (63). Since 3D-QALAS also uses an adiabatic inversion, we jointly estimated IE maps alongside $T_1$, $T_2$, and PD maps. It was recently demonstrated that solving for inversion efficiency in 3D-QALAS mapping increases the accuracy of the estimated $T_1$ and $T_2$ values when compared to the reference values in the NIST system phantom (64). While the original QALAS paper assumed complete inversion (i.e., IE = 1) and did not incorporate it in the dictionary generation or fitting, we assumed that the IE of an adiabatic inversion pulse is incomplete and generated the signal dictionary with IE that ranges between 0.5 to 1.

In the *in vivo* experiments, it was shown that the performance of the SSL-QALAS method was slightly lower for $T_2$ than for $T_1$ estimation. 3D-QALAS sequence uses a $T_2$ preparation pulse before the first measurement and an inversion pulse before the second acquisition, which indicates that $T_1$ relaxation is dependent on the five measurements, whereas $T_2$ relaxation is dependent on the first two measurements. Thus, $T_2$ estimation using the dictionary matching or SSL-QALAS method, which jointly estimates $T_1$, $T_2$, PD, and IE maps from five QALAS measurements, might be more sensitive to noise or motion than $T_1$ estimation. One possible way for robust $T_2$ estimation from QALAS measurements is to reduce the number of unknown values to be estimated by pre-acquiring the IE values, for example, using the Bloch simulation of the adiabatic inversion pulse, which can yield the IE value for each $T_1$ and $T_2$ pair instead of incorporating it in the dictionary generation or fitting.

It is important to note that recent studies have utilized the 3D-QALAS sequence for quantitative imaging and used a commercial software tool (SyMRI), which enabled rapid (< 1 min) reconstruction of multiparametric maps. While these studies have shown high reproducibility and repeatability of the acquired maps, the proprietary reconstruction software does not incorporate an external $B_1^+$ map for field inhomogeneity correction and does not solve for/incorporate IE.

Finally, the proposed method can be generalized and applied to other quantitative MR methods, such as EPTI and MRF, which usually require a large dictionary to store the time-series signals. The proposed SSL scheme for multiparametric mapping could be helpful in reducing the reconstruction time of the multiparametric maps.

## CONCLUSION

We proposed a rapid SSL-based quantitative mapping method called SSL-QALAS for rapid reconstruction of multiparametric maps, including $T_1$, $T_2$, PD, and IE maps from 3D-QALAS measurements, while obviating the need for an external dictionary and labeled ground-truth training data.

## ACKNOWLEDGMENTS

This work was supported by research grants NIH R01 EB032378, R01 EB028797, R03 EB031175, U01 EB025162, P41 EB030006, U01 EB026996, U01 DA055353 and the NVIDIA Corporation for computing support.

**Figure Legends**

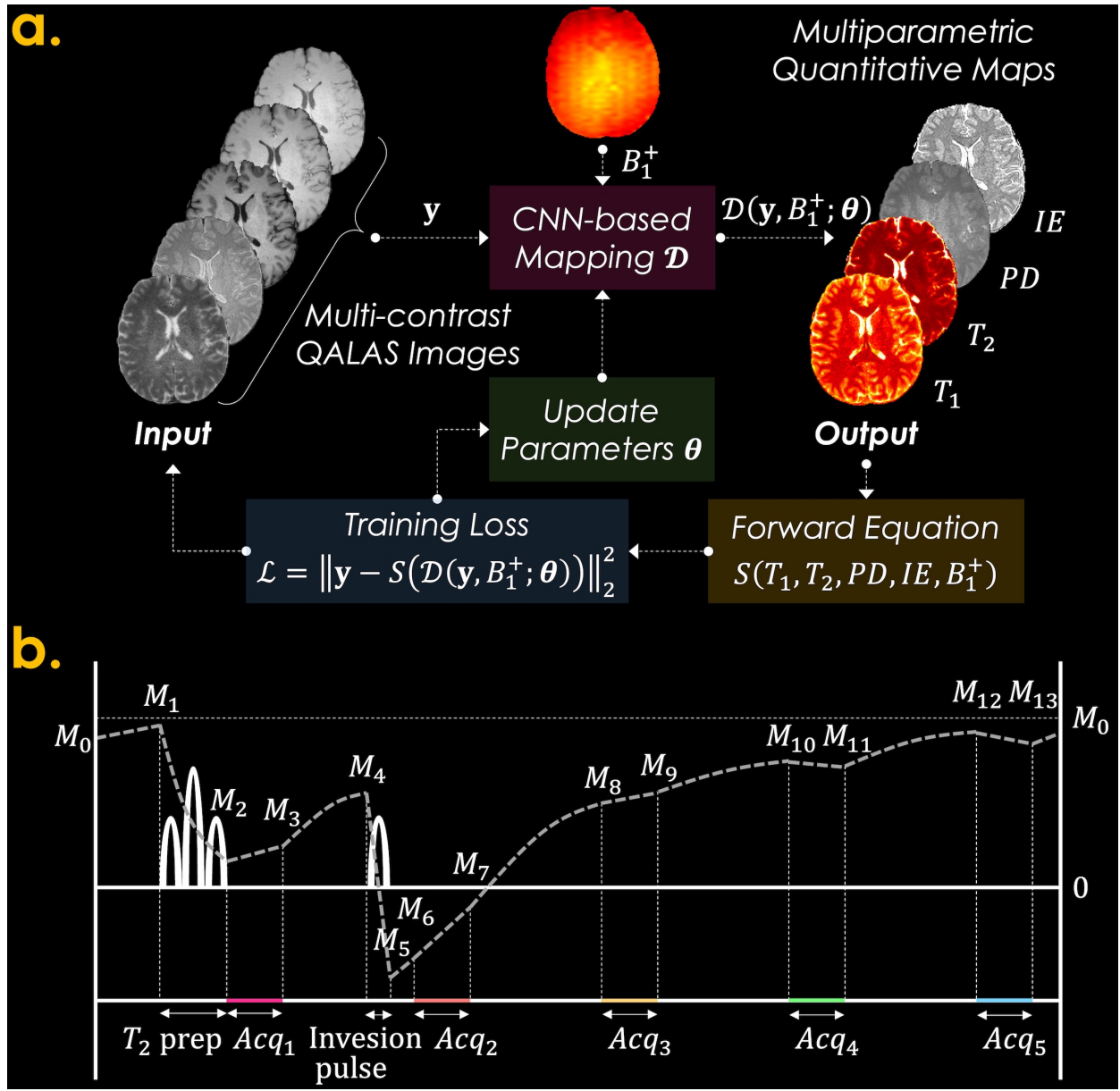

Figure. 1. (a) Overall flowchart of the proposed self-supervised learning (SSL) based method (i.e., SSL-QALAS) for multiparametric quantitative mapping using 3D-quantification using an interleaved Look-Locker acquisition sequence with $T_2$ preparation pulse (3D-QALAS). The five acquired QALAS contrast images along with $B_1^+$ map are fed into the convolutional neural network (CNN) model as the input, and it estimates four quantitative maps including $T_1$, $T_2$, proton density (PD), and inversion efficiency (IE) maps as the output. The model is trained by calculating the loss between the acquired images and synthetic images, which are generated by feeding the output maps into the forward signal model. (b) Sequence diagram of 3D-QALAS where it has a $T_2$ preparation pulse before the first data acquisition and an inversion pulse before the second data acquisition.

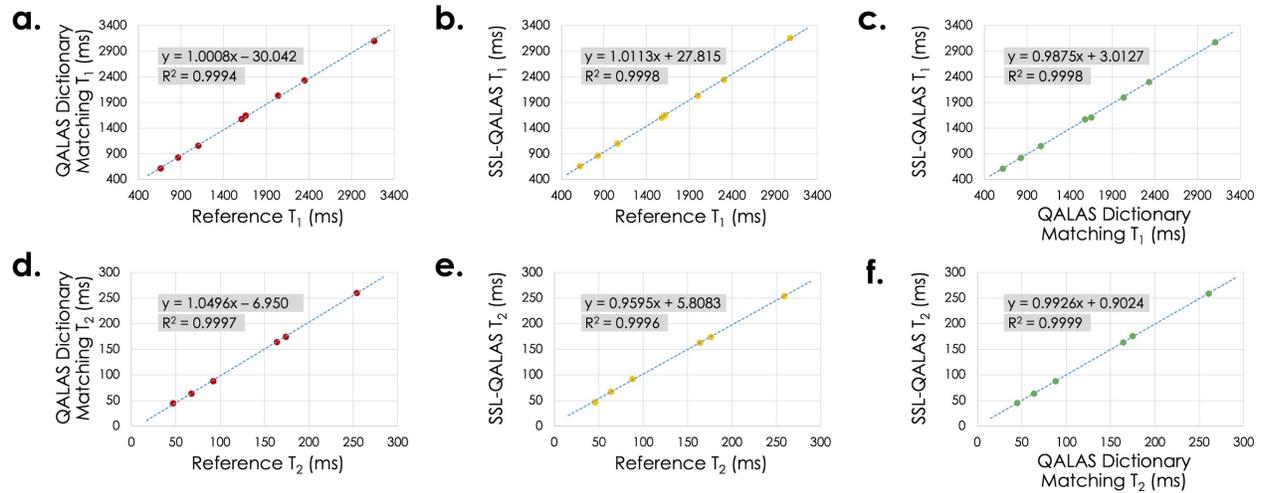

Figure. 2. Quantitative analyses of 3D-QALAS $T_1$ and $T_2$ maps using an ISMRM/NIST system phantom. The reference $T_1$ and $T_2$ maps were acquired using an IR-FSE and SE-FSE, respectively. Comparisons of $T_1$ and $T_2$ values (a, d) between the dictionary matching and reference methods, (b, e) between the proposed SSL-QALAS and reference methods, and (c, f) between the dictionary matching and the SSL-QALAS methods. 3D-QALAS: 3D-quantification using an interleaved Look-Locker acquisition sequence with $T_2$ preparation pulse; ISMRM/NIST: International Society for Magnetic Resonance in Medicine and National Institute of Standards and Technology; IR-FSE: inversion-recovery fast-spin-echo; SE-FSE: single-echo fast-spin-echo.

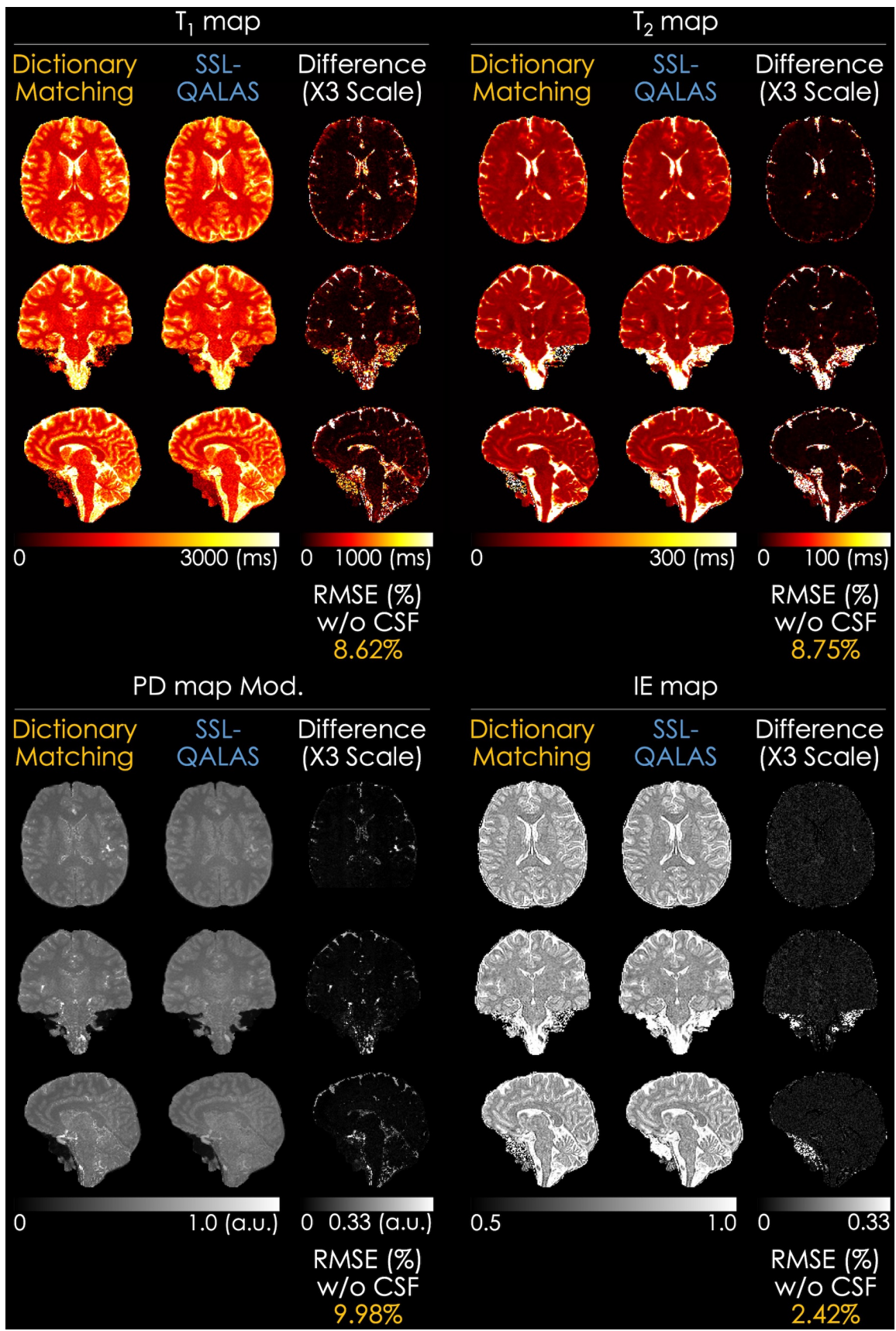

Figure. 3. Reconstructed $T_1$, $T_2$, proton density (PD), and inversion efficiency (IE) maps using the dictionary matching and proposed scan-specific SSL-QALAS methods in axial, coronal, and sagittal directions. The SSL-QALAS method shows comparable quantitative maps with the dictionary matching method. Root mean square error (RMSE) was calculated between the dictionary matching and the scan-specific SSL-QALAS methods in brain parenchyma regions.

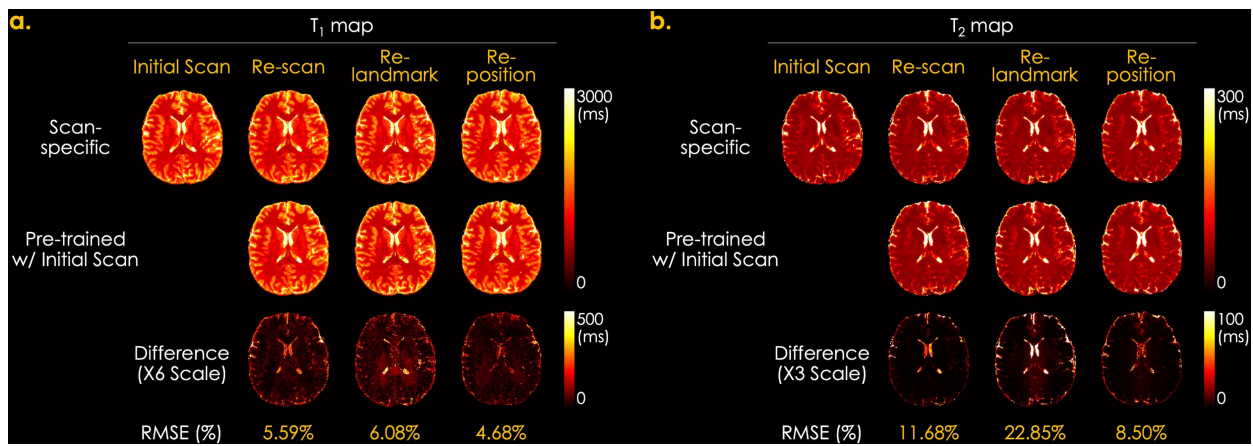

Figure. 4. Reconstructed (a) $T_1$ and (b) $T_2$ maps using the proposed SSL-QALAS method with 4 different acquisitions (i.e., initial, re-scan, re-landmark, and re-position) scanned from the same subject. The $T_1$ and $T_2$ maps in the first row were reconstructed using a scan-specific trained model, while those in the second row were reconstructed using the pre-trained model with the initial scan data. The difference images show the difference between the images of the scan-specific model and the ones of the pre-trained model. Root mean square error (RMSE) was calculated between the scan-specific model and the pre-trained model for each scan data.

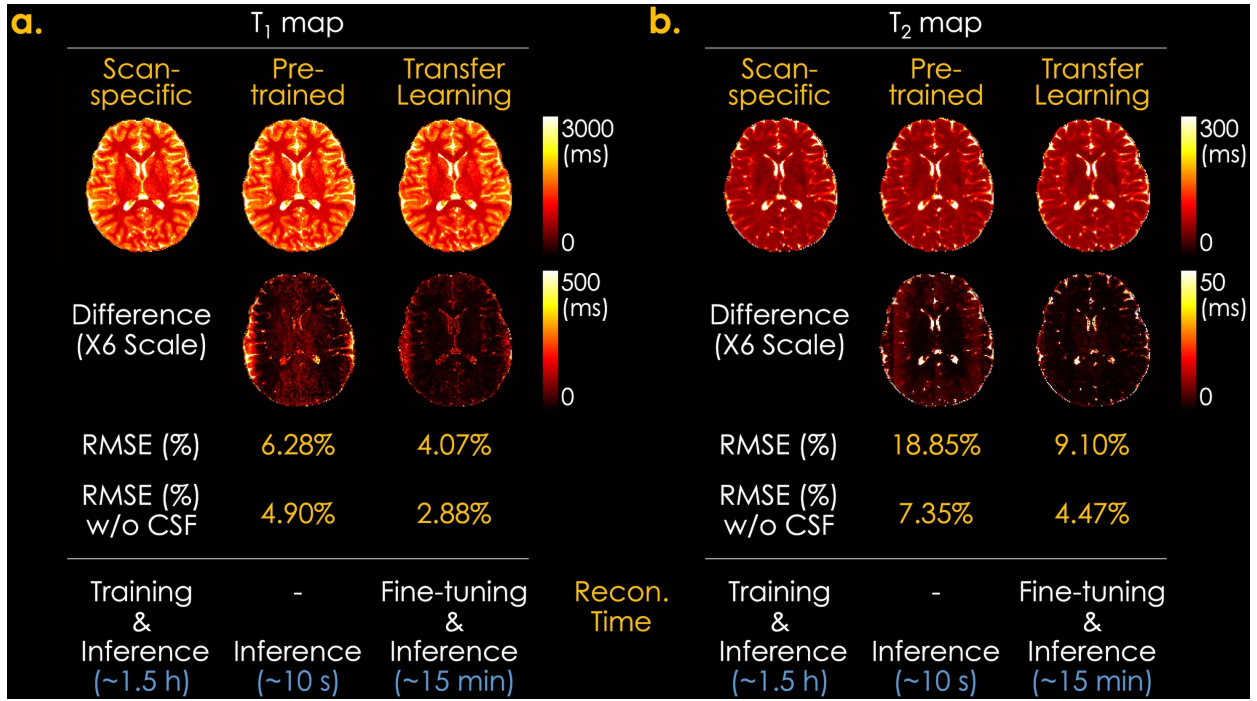

Figure. 5. Reconstructed (a) $T_1$ and (b) $T_2$ maps using the proposed SSL-QALAS method with 3 different models (i.e., scan-specific, pre-trained, and transfer learning models). The pre-trained model was trained with the other subject's data, and the transfer learning model was initially trained with the other subject's data and fine-tuned with the target subject's data. The difference images show the difference between the reconstructed images of the scan-specific model and the ones of the pre-trained or transfer learning models. The reconstruction for each model takes about 1.5 h (scan-specific: training and inference), 10 s (pre-trained: inference only), and 15 min (transfer learning: fine-tuning and inference), respectively. Root mean square error (RMSE) was calculated between the scan-specific model and the pre-trained or transfer learning model.

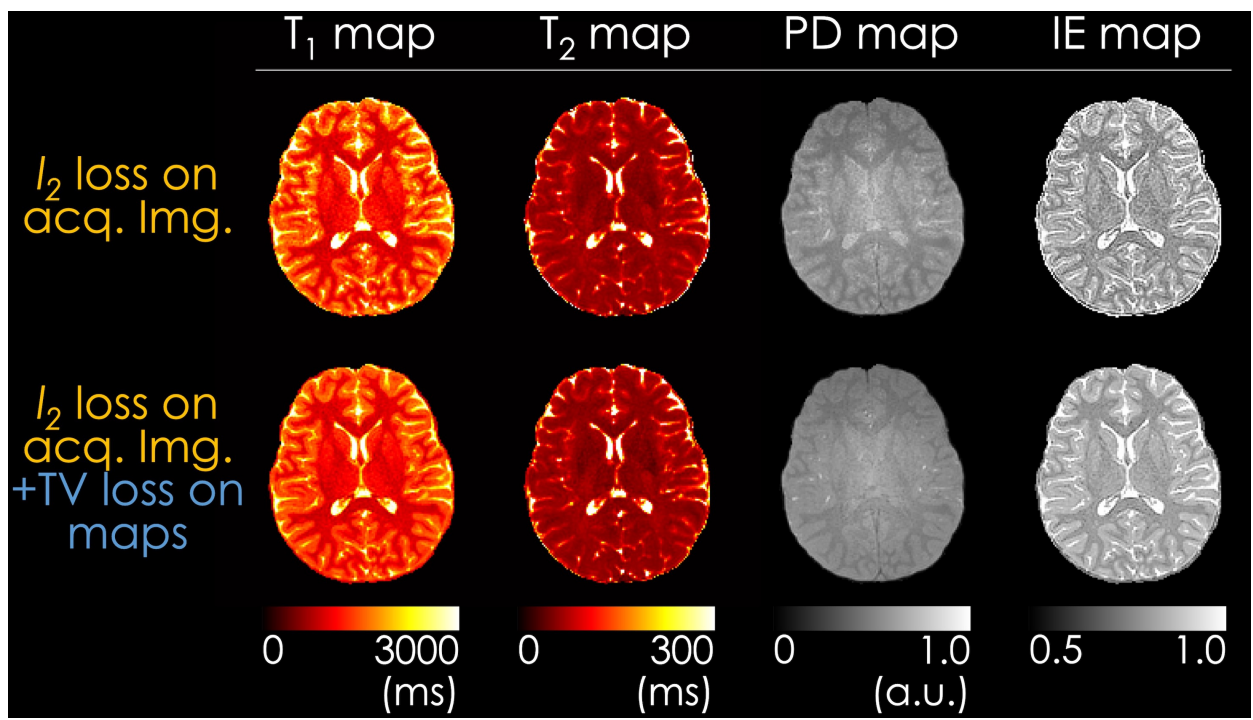

Figure. 6. Reconstructed $T_1$, $T_2$, proton density (PD), and inversion efficiency (IE) maps using the proposed scan-specific SSL-QALAS model. The first row shows the original SSL-QALAS model, which used $l_2$ loss on the acquired images, whereas the second row shows the modified model, which additionally used total variation (TV) loss on the quantitative maps.

**Supporting Information Legends**

Supporting Information Table. S1. MRI scan parameters of 3D-QALAS and turbo-FLASH sequences.

Supporting Information Table S2. MRI scan parameters of IR-FSE and SE-FSE sequences for ISMRM/NIST system phantom experiment.

Supporting Information Table S3. Quantitative analysis of pre-trained and transfer learning SSL-QALAS models compared with a scan-specific SSL-QALAS model.

Supporting Information Table S4. Quantitative analysis of SSL-QALAS models compared with a dictionary matching method.

Supporting Information Figure S1. Architecture of the convolutional neural network (CNN) implemented in SSL-QALAS method.

Supporting Information Figure S2. Quantitative analyses of 3D-QALAS $T_1$ maps using an ISMRM/NIST system phantom. The reference $T_1$ maps were acquired using an IR-FSE sequence. Comparisons of $T_1$ values (a, b) between the dictionary matching and reference methods using the initial scan and re-scan data, respectively, (c) between the initial scan and re-scan data using the dictionary matching method, (d, e) between the SSL-QALAS and reference methods using the initial scan and re-scan data, respectively, (f) between the initial scan and re-scan data using the SSL-QALAS method, (g) between the pre-trained SSL-QALAS and reference methods using the re-scan data, (h) between the initial scan and re-scan data using the pre-trained SSL-QALAS method, and (i) between the scan-specific and pre-trained SSL-QALAS methods using the re-scan data. 3D-QALAS: 3D-quantification using an interleaved Look-Locker acquisition sequence with $T_2$ preparation pulse; ISMRM/NIST: International Society for Magnetic Resonance in Medicine and National Institute of Standards and Technology; IR-FSE: inversion-recovery fast-spin-echo.

Supporting Information Figure S3. Quantitative analyses of 3D-QALAS $T_2$ maps using an ISMRM/NIST system phantom. The reference $T_2$ maps were acquired using a SE-FSE sequence.

Comparisons of $T_2$ values (a, b) between the dictionary matching and reference methods using the initial scan and re-scan data, respectively, (c) between the initial scan and re-scan data using the dictionary matching method, (d, e) between the SSL-QALAS and reference methods using the initial scan and re-scan data, respectively, (f) between the initial scan and re-scan data using the SSL-QALAS method, (g) between the pre-trained SSL-QALAS and reference methods using the re-scan data, (h) between the initial scan and re-scan data using the pre-trained SSL-QALAS method, and (i) between the scan-specific and pre-trained SSL-QALAS methods using the re-scan data. 3D-QALAS: 3D-quantification using an interleaved Look-Locker acquisition sequence with $T_2$ preparation pulse; ISMRM/NIST: International Society for Magnetic Resonance in Medicine and National Institute of Standards and Technology; SE-FSE: single-echo fast-spin-echo.

Supporting Information Figure S4. Difference images between the reconstructed (a) $T_1$ and (b) $T_2$ maps using the proposed SSL-QALAS and dictionary matching methods with 4 different acquisitions (i.e., initial, re-scan, re-landmark, and re-position) scanned from the same subject. The $T_1$ and $T_2$ maps in the first row were reconstructed using a scan-specific trained model, while those in the second row were reconstructed using the pre-trained model with the initial scan data. Root mean square error (RMSE) was calculated between the SSL-QALAS and dictionary matching methods for each scan data except for some cerebrospinal fluid (CSF) regions.

Supporting Information Figure S5. Difference images between the reconstructed (a) $T_1$ and (b) $T_2$ maps of re-scan, re-landmark, and re-position, and those of initial scan using the proposed SSL-QALAS. Those 4 different acquisitions (i.e., initial, re-scan, re-landmark, and re-position) were scanned from the same subject. The $T_1$ and $T_2$ maps in the first row were reconstructed using a scan-specific trained model, while those in the second row were reconstructed using the pre-trained model with the initial scan data. Root mean square error (RMSE) was calculated between the re-scan, re-landmark, and re-position, and the initial scan using the proposed SSL-QALAS for each scan data except for some cerebrospinal fluid (CSF) regions.

Supporting Information Figure S6. Reconstructed (a) proton density (PD) and (b) inversion efficiency (IE) maps using the proposed SSL-QALAS method with 3 different models (i.e., scan-specific, pre-trained, and transfer learning models). The pre-trained model was trained with the other subject's data, and the transfer learning model was initially trained with the other subject's data and fine-tuned with the target subject's data. The difference images show the difference between the reconstructed images of the scan-specific model and the ones of the pre-trained or transfer learning models. The reconstruction for each model takes about 1.5 h (scan-specific: training and inference), 10 s (pre-trained: inference only), and 15 min (transfer learning: fine-tuning and inference), respectively. Root mean square error (RMSE) was calculated between the scan-specific model and the pre-trained or transfer learning model.

Supporting Information Figure S7. Reconstructed (a) $T_1$ and (b) $T_2$ maps using the proposed SSL-QALAS method with 3 different models (i.e., scan-specific, pre-trained, and transfer learning models). The pre-trained model was trained with the other subject's data, and the transfer learning model was initially trained with the other subject's data and fine-tuned with the target subject's data. The difference images show the difference between the reconstructed images of the SSL-QALAS method and the ones of the dictionary matching method. The reconstruction for each model takes about 1.5 h (scan-specific: training and inference), 10 s (pre-trained: inference only), and 15 min (transfer learning: fine-tuning and inference), respectively. Root mean square error (RMSE) was calculated between the SSL-QALAS method and the dictionary matching method.

Supporting Information Figure S8. Reconstructed (a) proton density (PD) and (b) inversion efficiency (IE) maps using the proposed SSL-QALAS method with 3 different models (i.e., scan-specific, pre-trained, and transfer learning models). The pre-trained model was trained with the other subject's data, and the transfer learning model was initially trained with the other subject's data and fine-tuned with the target subject's data. The difference images show the difference between the reconstructed images of the SSL-QALAS method and the ones of the dictionary matching method. The reconstruction for each model takes about 1.5 h (scan-specific: training and inference), 10 s (pre-trained: inference only), and 15 min (transfer learning: fine-tuning and inference), respectively. Root mean square error (RMSE) was calculated between the SSL-QALAS method and the dictionary matching method.

Supporting Information Figure S9. Reconstructed (a) $T_1$ and (b) $T_2$ maps using the proposed SSL-QALAS method with 3 different models (i.e., scan-specific, pre-trained, and transfer learning models), which used 33 convolutional layers. The pre-trained model was trained with the other subject's data, and the transfer learning model was initially trained with the other subject's data and fine-tuned with the target subject's data. The difference images show the difference between the reconstructed images of the scan-specific model and the ones of the pre-trained or transfer learning models. The reconstruction for each model takes about 1.5 h (scan-specific: training and inference), 10 s (pre-trained: inference only), and 15 min (transfer learning: fine-tuning and inference), respectively. Root mean square error (RMSE) was calculated between the scan-specific model and the pre-trained or transfer learning model.

Supporting Information Figure S10. Reconstructed (a) proton density (PD) and (b) inversion efficiency (IE) maps using the proposed SSL-QALAS method with 3 different models (i.e., scan-specific, pre-trained, and transfer learning models), which used 3×3 convolutional layers. The pre-trained model was trained with the other subject's data, and the transfer learning model was initially trained with the other subject's data and fine-tuned with the target subject's data. The difference images show the difference between the reconstructed images of the scan-specific model and the ones of the pre-trained or transfer learning models. The reconstruction for each model takes about 1.5 h (scan-specific: training and inference), 10 s (pre-trained: inference only), and 15 min (transfer learning: fine-tuning and inference), respectively. Root mean square error (RMSE) was calculated between the scan-specific model and the pre-trained or transfer learning model.

Supporting Information Figure S11. Reconstructed (a) $T_1$ and (b) $T_2$ maps using the proposed SSL-QALAS method with 3 different models (i.e., scan-specific, pre-trained, and transfer learning models). The pre-trained model was trained with the other subject's data, which has 1.15 mm$^3$ isotropic resolution, and the transfer learning model was initially trained with the other subject's data and fine-tuned with the target subject's data, which has 1.3 mm$^3$ isotropic resolution. The difference images show the difference between the reconstructed images of the scan-specific model and the ones of the pre-trained or transfer learning models. The reconstruction for each model takes about 1.5 h (scan-specific: training and inference), 10 s (pre-trained: inference only), and 15 min (transfer learning: fine-tuning and inference), respectively. Root mean square error (RMSE) was calculated between the scan-specific model and the pre-trained or transfer learning model.

Supporting Information Figure S12. Reconstructed (a) $T_1$ and (b) $T_2$ maps using the proposed SSL-QALAS method with 3 different models (i.e., scan-specific, pre-trained, and transfer learning models). The pre-trained model was trained with the other subject's data, which has 1.3 mm$^3$ isotropic resolution, and the transfer learning model was initially trained with the other subject's data and fine-tuned with the target subject's data, which has 1.15 mm$^3$ isotropic resolution. The difference images show the difference between the reconstructed images of the scan-specific model and the ones of the pre-trained or transfer learning models. The reconstruction for each model takes about 1.5 h (scan-specific: training and inference), 10 s (pre-trained: inference only), and 15 min (transfer learning: fine-tuning and inference), respectively. Root mean square error (RMSE) was calculated between the scan-specific model and the pre-trained or transfer learning model.

**Supporting Information Data**

Supporting Information Table S1. MRI scan parameters of 3D-QALAS and turbo-FLASH sequences.

|  | ISMRM/NIST Phantom | In vivo Experiment #1 (Repeatability) | In vivo Experiment #2 (Generalizability) |
|---|---|---|---|
| **3D-QALAS** | | | |
| FOV | 224 × 256 × 216 mm$^3$ | 228 × 228 × 208 mm$^3$ | 240 × 240 × 202 mm$^3$ |
| Matrix Size | 224 × 256 × 72 | 176 × 176 × 160 | 208 × 208 × 176 |
| BW | 340 Hz/pixel | 320 Hz/pixel | 330 Hz/pixel |
| Echo Spacing | 5.8 ms | 5.7 ms | 5.76 ms |
| Turbo Factor | 127 | 128 | 128 |
| TR | 4.5 s | 4.5 s | 4.5 s |
| TE | 2.29 ms | 2.3 ms | 2.29 ms |
| Acceleration | 1 | 2 | 2 |
| Scan Time | 7 min 26 s | 6 min 34 s | 8 min 24 s |
| **Turbo-FLASH $B_1^+$** | | | |
| FOV | 224 × 256 mm$^2$ | 228 × 228 mm$^2$ | 240 × 240 mm$^2$ |
| Matrix Size | 84 × 96 | 64 × 64 | 64 × 64 |
| Number of Slices | 60 | 58 | 28 |
| Slice Thickness | 3 mm | 3 mm | 3 mm |
| BW | 490 Hz/pixel | 490 Hz/pixel | 490 Hz/pixel |
| TR | 16.16 s | 12.68 s | 20 s |
| TE | 2.66 ms | 2.66 ms | 2.66 ms |
| Acceleration | 2 | 2 | 2 |
| Scan Time | 34 s | 27 s | 21 s |

3D-QALAS: 3D-quantification using an interleaved Look-Locker acquisition sequence with $T_2$ preparation pulse; turbo-FLASH: turbo-fast low-angle shot sequence for $B_1^+$ mapping

Supporting Information Table S2. MRI scan parameters of IR-FSE and SE-FSE sequences for ISMRM/NIST system phantom experiment.

|  | **IR-FSE** | **SE-FSE** |
|---|---|---|
| FOV | 224 × 256 mm$^2$ | 224 × 256 mm$^2$ |
| Matrix Size | 224 × 256 | 224 × 256 |
| Slice Thickness | 3 mm | 3 mm |
| BW | 227 Hz/pixel | 227 Hz/pixel |
| TR | 8 s | 2 s |
| TE | 9.2 ms | 10.0 ms |
| TI | [10, 20, 30, 40, 50, 60, 70, 80, 90, 100, 120, 150, 200, 250, 300] ms | [25, 50, 75, 100, 150, 200, 250, 500, 750, 1000, 1250, 1500, 1750, 2000, 2500, 3000, 5000] ms |
| Turbo Factor | 21 | 124 |
| Acceleration | 2 | 2 |
| Scan Time | 2 min 58 s | 4 min 12 s |

ISMRM/NIST: International Society for Magnetic Resonance in Medicine and National Institute of Standards and Technology; IR-FSE: inversion-recovery fast-spin-echo; SE-FSE: single-echo fast-spin-echo

Supporting Information Table S3. Quantitative analysis of pre-trained and transfer learning SSL-QALAS models compared with a scan-specific SSL-QALAS model.

| %RMSE w/o CSF | $T_1$ | $T_2$ | PD | IE |
|---|---|---|---|---|
| **Pre-trained** | 4.50 ± 0.84 | 9.74 ± 3.38 | 3.41 ± 0.54 | 1.20 ± 0.13 |
| **Transfer Learning** | 2.84 ± 0.39 | 5.51 ± 1.39 | 2.14 ± 0.34 | 1.02 ± 0.15 |

Note: RMSE was calculated between the scan-specific model and the pre-trained or transfer learning model. RMSE: root mean square error; CSF: cerebrospinal fluid

Supporting Information Table S4. Quantitative analysis of SSL-QALAS models compared with a dictionary matching method.

| %RMSE w/o CSF | $T_1$ | $T_2$ | PD | IE |
|---|---|---|---|---|
| **Scan-specific** | 9.05 ± 0.98 | 14.19 ± 3.35 | 13.73 ± 1.66 | 1.14 ± 0.06 |
| **Pre-trained** | 10.12 ± 1.29 | 17.15 ± 3.36 | 12.71 ± 1.73 | 1.34 ± 0.08 |
| **Transfer Learning** | 8.62 ± 0.83 | 13.80 ± 2.56 | 12.78 ± 1.67 | 1.15 ± 0.11 |

Note: RMSE was calculated between the SSL-QALAS models and dictionary matching method.

RMSE: root mean square error; CSF: cerebrospinal fluid

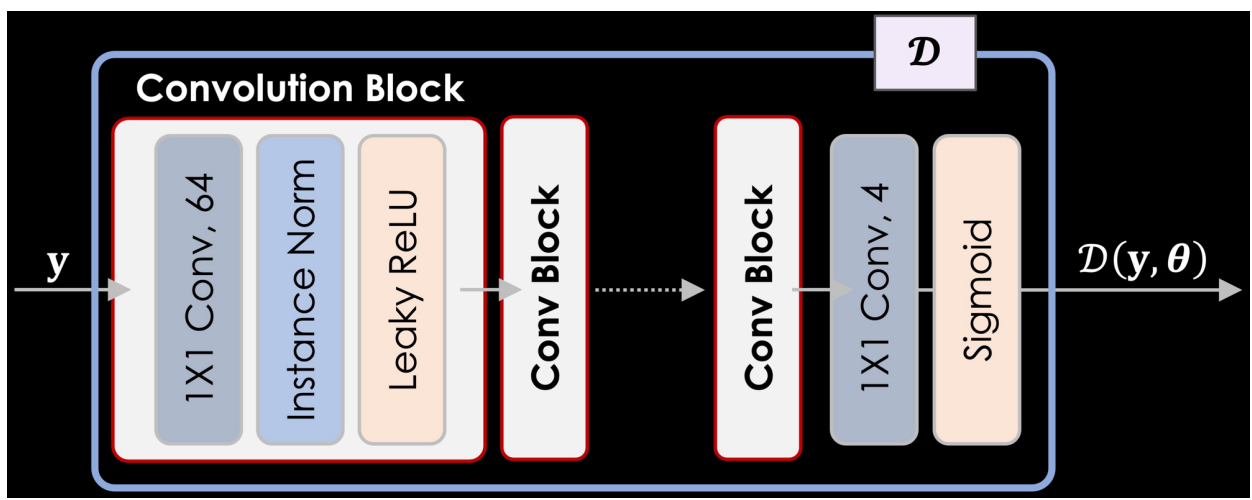

Supporting Information Figure S1. Architecture of the convolutional neural network (CNN) implemented in SSL-QALAS method.

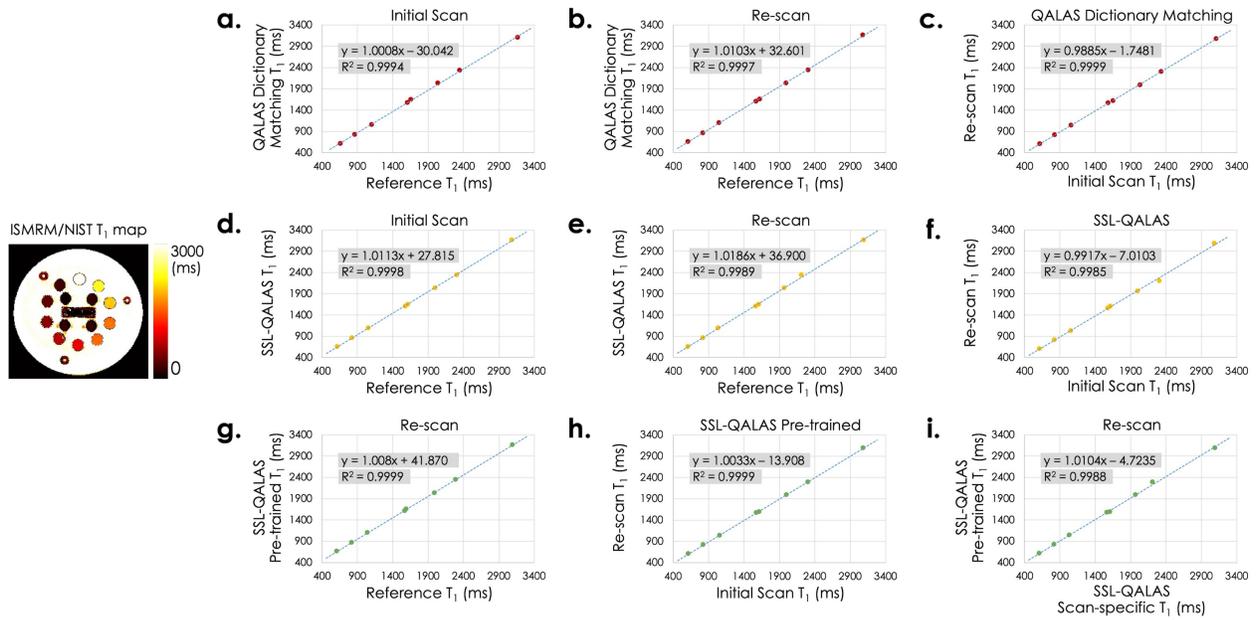

Supporting Information Figure S2. Quantitative analyses of 3D-QALAS $T_1$ maps using an ISMRM/NIST system phantom. The reference $T_1$ maps were acquired using an IR-FSE sequence. Comparisons of $T_1$ values (a, b) between the dictionary matching and reference methods using the initial scan and re-scan data, respectively, (c) between the initial scan and re-scan data using the dictionary matching method, (d, e) between the SSL-QALAS and reference methods using the initial scan and re-scan data, respectively, (f) between the initial scan and re-scan data using the SSL-QALAS method, (g) between the pre-trained SSL-QALAS and reference methods using the re-scan data, (h) between the initial scan and re-scan data using the pre-trained SSL-QALAS method, and (i) between the scan-specific and pre-trained SSL-QALAS methods using the re-scan data. 3D-QALAS: 3D-quantification using an interleaved Look-Locker acquisition sequence with $T_2$ preparation pulse; ISMRM/NIST: International Society for Magnetic Resonance in Medicine and National Institute of Standards and Technology; IR-FSE: inversion-recovery fast-spin-echo.

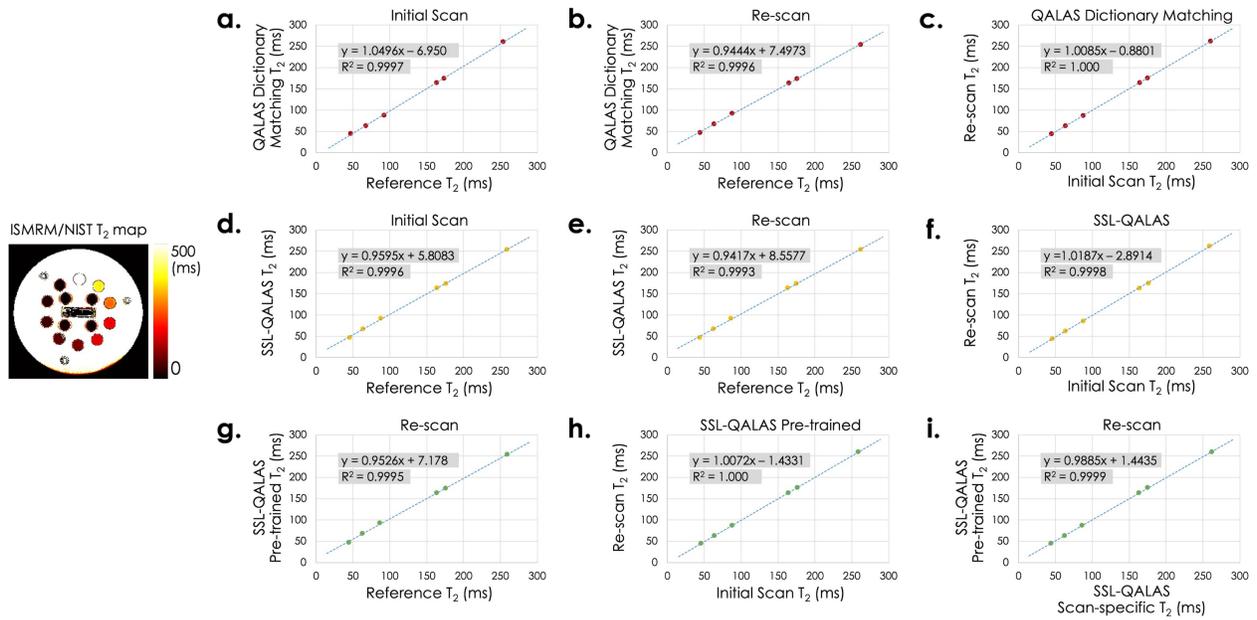

Supporting Information Figure S3. Quantitative analyses of 3D-QALAS $T_2$ maps using an ISMRM/NIST system phantom. The reference $T_2$ maps were acquired using a SE-FSE sequence. Comparisons of $T_2$ values (a, b) between the dictionary matching and reference methods using the initial scan and re-scan data, respectively, (c) between the initial scan and re-scan data using the dictionary matching method, (d, e) between the SSL-QALAS and reference methods using the initial scan and re-scan data, respectively, (f) between the initial scan and re-scan data using the SSL-QALAS method, (g) between the pre-trained SSL-QALAS and reference methods using the re-scan data, (h) between the initial scan and re-scan data using the pre-trained SSL-QALAS method, and (i) between the scan-specific and pre-trained SSL-QALAS methods using the re-scan data. 3D-QALAS: 3D-quantification using an interleaved Look-Locker acquisition sequence with $T_2$ preparation pulse; ISMRM/NIST: International Society for Magnetic Resonance in Medicine and National Institute of Standards and Technology; SE-FSE: single-echo fast-spin-echo.

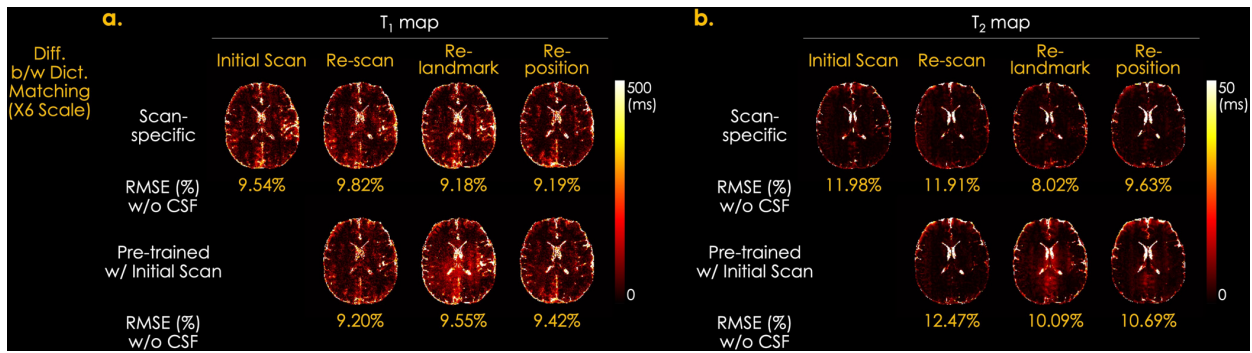

Supporting Information Figure S4. Difference images between the reconstructed (a) $T_1$ and (b) $T_2$ maps using the proposed SSL-QALAS and dictionary matching methods with 4 different acquisitions (i.e., initial, re-scan, re-landmark, and re-position) scanned from the same subject. The $T_1$ and $T_2$ maps in the first row were reconstructed using a scan-specific trained model, while those in the second row were reconstructed using the pre-trained model with the initial scan data. Root mean square error (RMSE) was calculated between the SSL-QALAS and dictionary matching methods for each scan data except for some cerebrospinal fluid (CSF) regions.

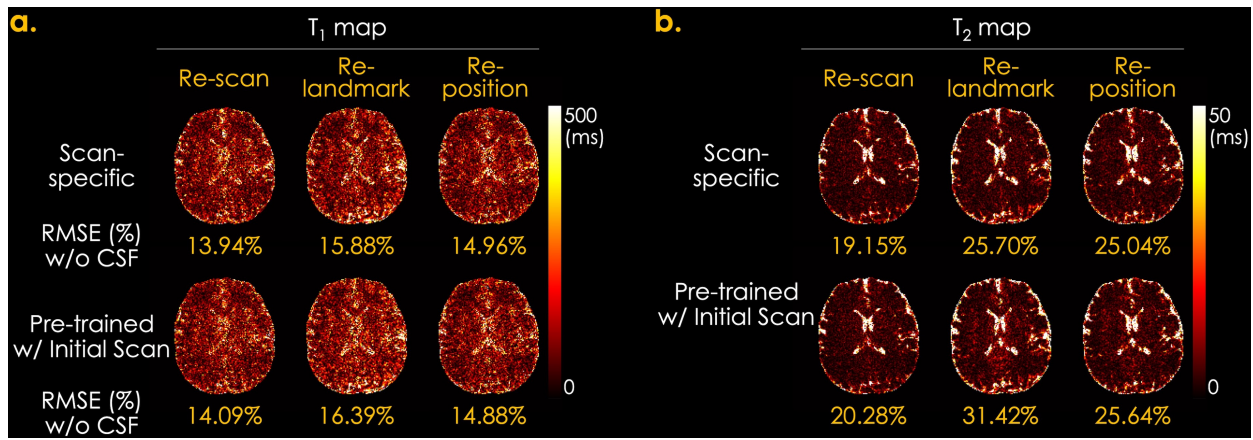

Supporting Information Figure S5. Difference images between the reconstructed (a) $T_1$ and (b) $T_2$ maps of re-scan, re-landmark, and re-position, and those of initial scan using the proposed SSL-QALAS. Those 4 different acquisitions (i.e., initial, re-scan, re-landmark, and re-position) were scanned from the same subject. The $T_1$ and $T_2$ maps in the first row were reconstructed using a scan-specific trained model, while those in the second row were reconstructed using the pre-trained model with the initial scan data. Root mean square error (RMSE) was calculated between the re-scan, re-landmark, and re-position, and the initial scan using the proposed SSL-QALAS for each scan data except for some cerebrospinal fluid (CSF) regions.

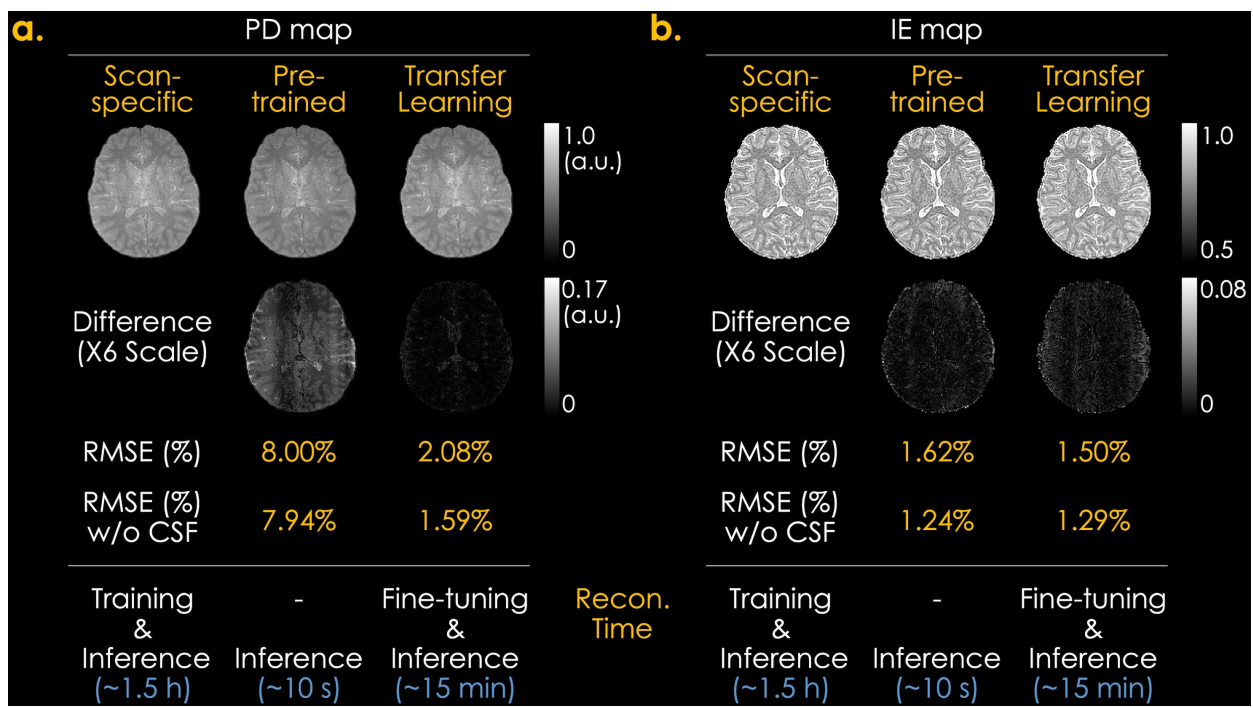

Supporting Information Figure S6. Reconstructed (a) proton density (PD) and (b) inversion efficiency (IE) maps using the proposed SSL-QALAS method with 3 different models (i.e., scan-specific, pre-trained, and transfer learning models). The pre-trained model was trained with the other subject's data, and the transfer learning model was initially trained with the other subject's data and fine-tuned with the target subject's data. The difference images show the difference between the reconstructed images of the scan-specific model and the ones of the pre-trained or transfer learning models. The reconstruction for each model takes about 1.5 h (scan-specific: training and inference), 10 s (pre-trained: inference only), and 15 min (transfer learning: fine-tuning and inference), respectively. Root mean square error (RMSE) was calculated between the scan-specific model and the pre-trained or transfer learning model.

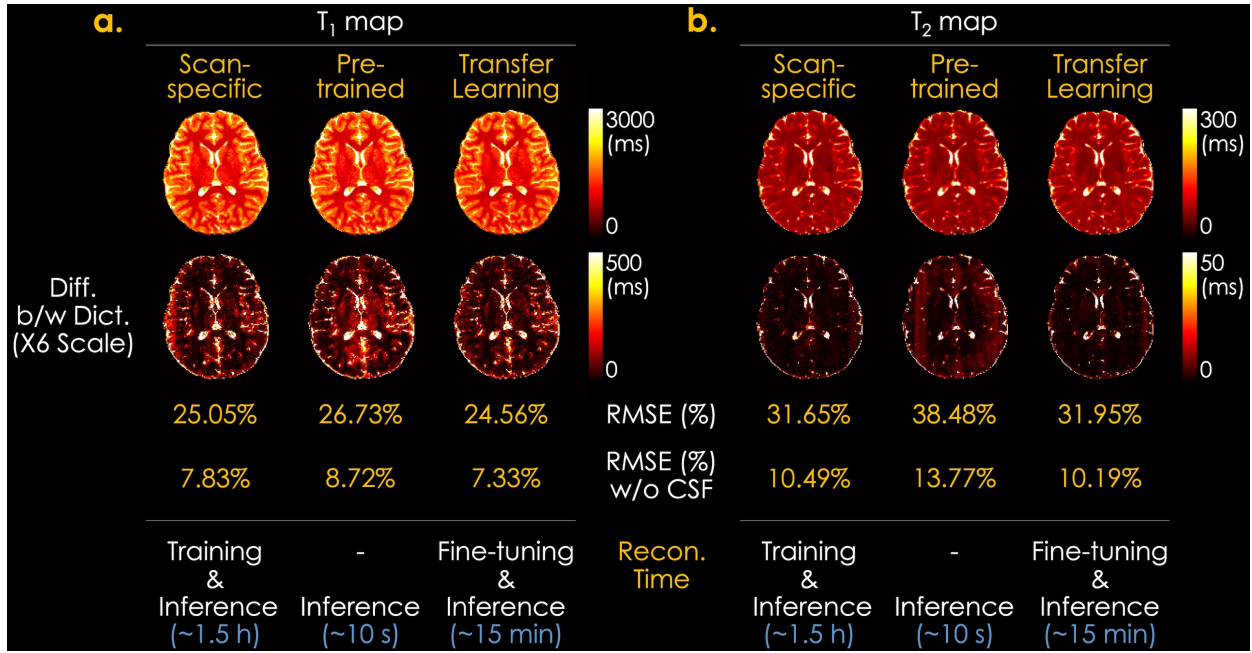

Supporting Information Figure S7. Reconstructed (a) $T_1$ and (b) $T_2$ maps using the proposed SSL-QALAS method with 3 different models (i.e., scan-specific, pre-trained, and transfer learning models). The pre-trained model was trained with the other subject's data, and the transfer learning model was initially trained with the other subject's data and fine-tuned with the target subject's data. The difference images show the difference between the reconstructed images of the SSL-QALAS method and the ones of the dictionary matching method. The reconstruction for each model takes about 1.5 h (scan-specific: training and inference), 10 s (pre-trained: inference only), and 15 min (transfer learning: fine-tuning and inference), respectively. Root mean square error (RMSE) was calculated between the SSL-QALAS method and the dictionary matching method.

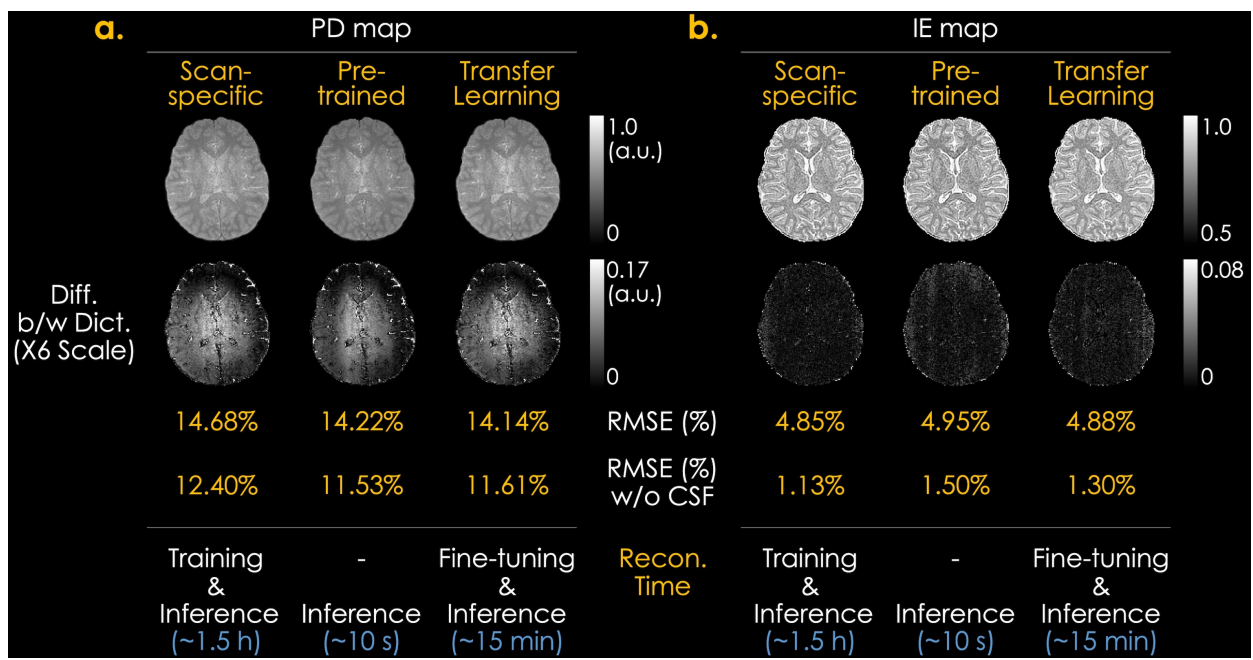

Supporting Information Figure S8. Reconstructed (a) proton density (PD) and (b) inversion efficiency (IE) maps using the proposed SSL-QALAS method with 3 different models (i.e., scan-specific, pre-trained, and transfer learning models). The pre-trained model was trained with the other subject's data, and the transfer learning model was initially trained with the other subject's data and fine-tuned with the target subject's data. The difference images show the difference between the reconstructed images of the SSL-QALAS method and the ones of the dictionary matching method. The reconstruction for each model takes about 1.5 h (scan-specific: training and inference), 10 s (pre-trained: inference only), and 15 min (transfer learning: fine-tuning and inference), respectively. Root mean square error (RMSE) was calculated between the SSL-QALAS method and the dictionary matching method.

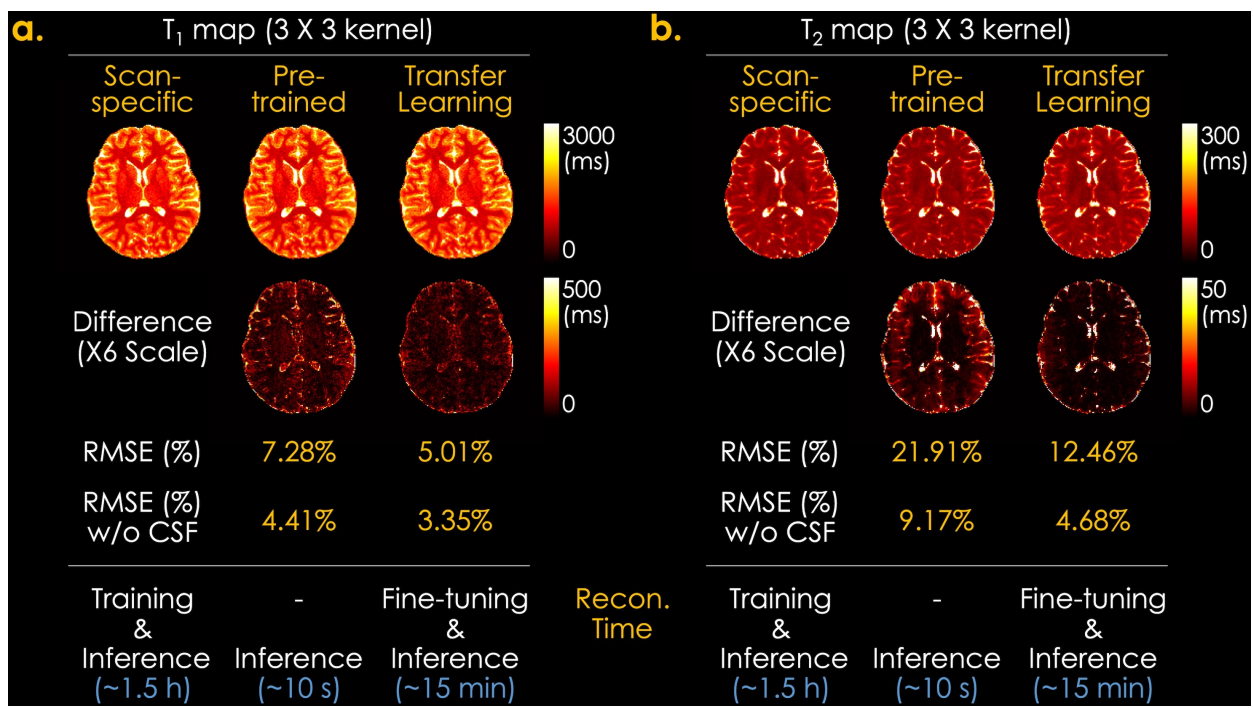

Supporting Information Figure S9. Reconstructed (a) $T_1$ and (b) $T_2$ maps using the proposed SSL-QALAS method with 3 different models (i.e., scan-specific, pre-trained, and transfer learning models), which used 3×3 convolutional layers. The pre-trained model was trained with the other subject's data, and the transfer learning model was initially trained with the other subject's data and fine-tuned with the target subject's data. The difference images show the difference between the reconstructed images of the scan-specific model and the ones of the pre-trained or transfer learning models. The reconstruction for each model takes about 1.5 h (scan-specific: training and inference), 10 s (pre-trained: inference only), and 15 min (transfer learning: fine-tuning and inference), respectively. Root mean square error (RMSE) was calculated between the scan-specific model and the pre-trained or transfer learning model.

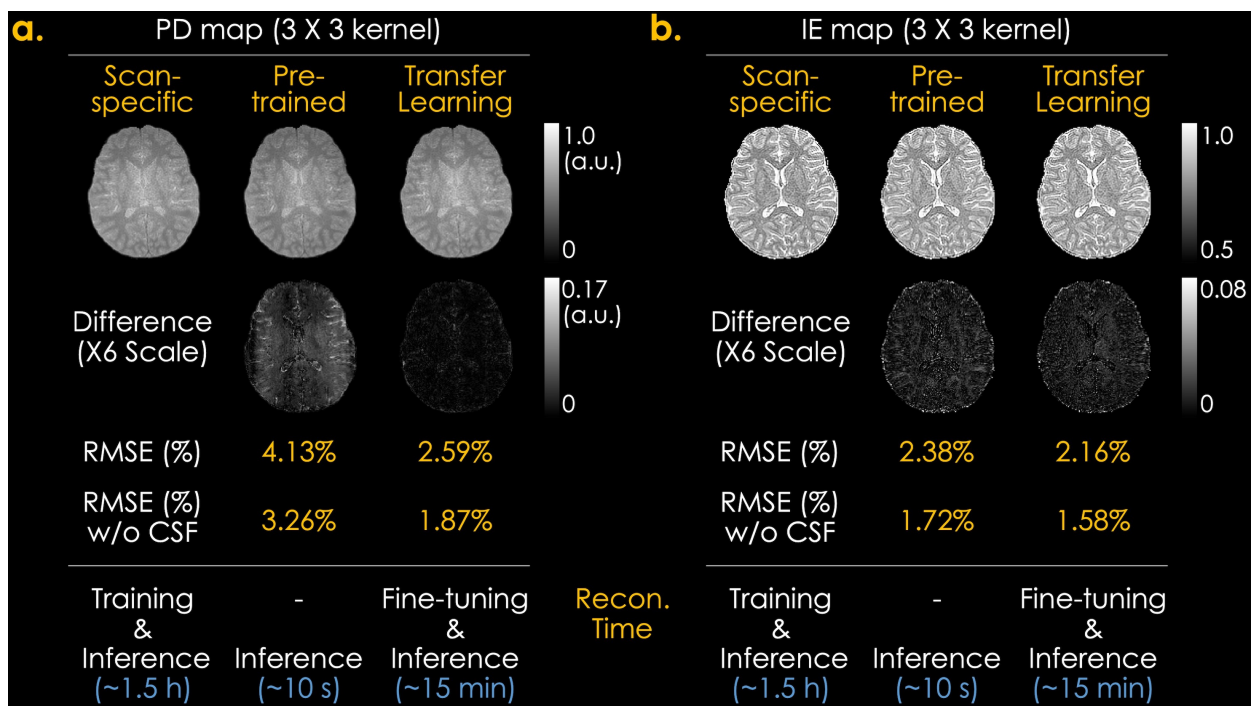

Supporting Information Figure S10. Reconstructed (a) proton density (PD) and (b) inversion efficiency (IE) maps using the proposed SSL-QALAS method with 3 different models (i.e., scan-specific, pre-trained, and transfer learning models), which used 3×3 convolutional layers. The pre-trained model was trained with the other subject's data, and the transfer learning model was initially trained with the other subject's data and fine-tuned with the target subject's data. The difference images show the difference between the reconstructed images of the scan-specific model and the ones of the pre-trained or transfer learning models. The reconstruction for each model takes about 1.5 h (scan-specific: training and inference), 10 s (pre-trained: inference only), and 15 min (transfer learning: fine-tuning and inference), respectively. Root mean square error (RMSE) was calculated between the scan-specific model and the pre-trained or transfer learning model.

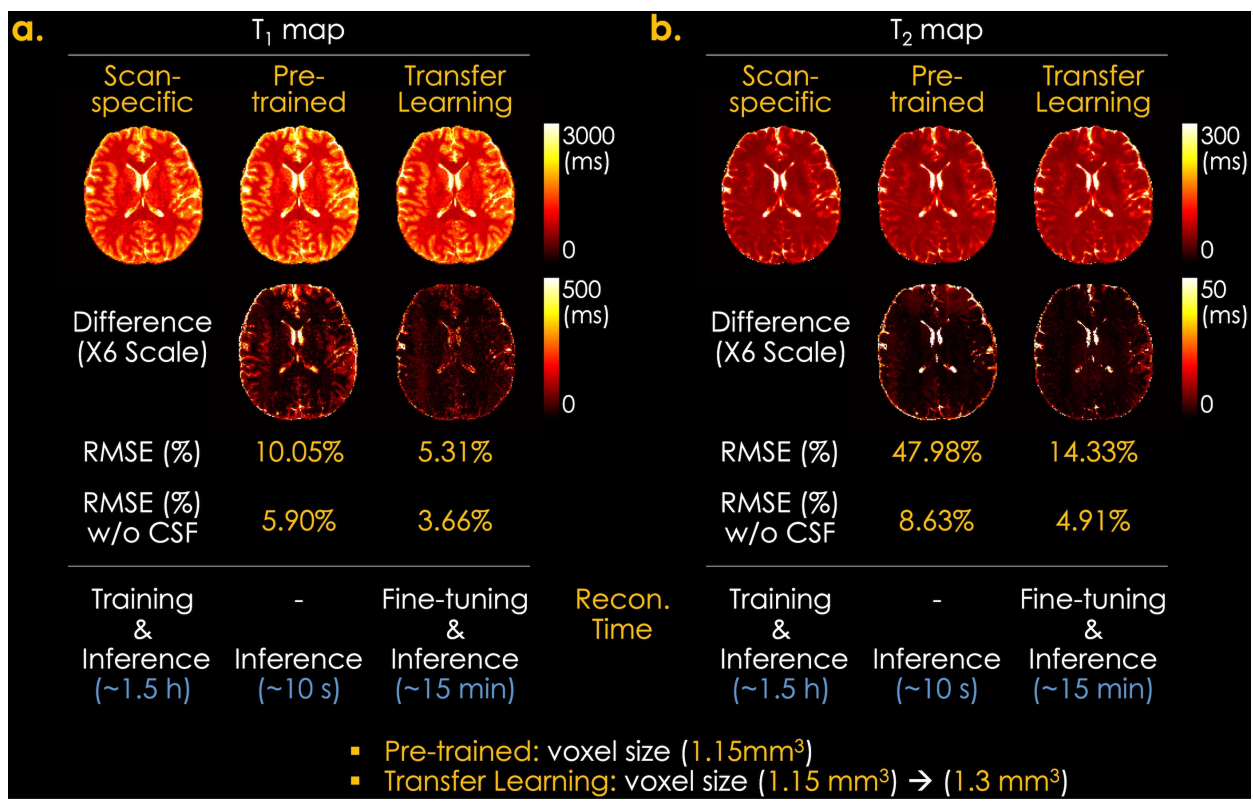

Supporting Information Figure S11. Reconstructed (a) $T_1$ and (b) $T_2$ maps using the proposed SSL-QALAS method with 3 different models (i.e., scan-specific, pre-trained, and transfer learning models). The pre-trained model was trained with the other subject's data, which has 1.15 mm$^3$ isotropic resolution, and the transfer learning model was initially trained with the other subject's data and fine-tuned with the target subject's data, which has 1.3 mm$^3$ isotropic resolution. The difference images show the difference between the reconstructed images of the scan-specific model and the ones of the pre-trained or transfer learning models. The reconstruction for each model takes about 1.5 h (scan-specific: training and inference), 10 s (pre-trained: inference only), and 15 min (transfer learning: fine-tuning and inference), respectively. Root mean square error (RMSE) was calculated between the scan-specific model and the pre-trained or transfer learning model.

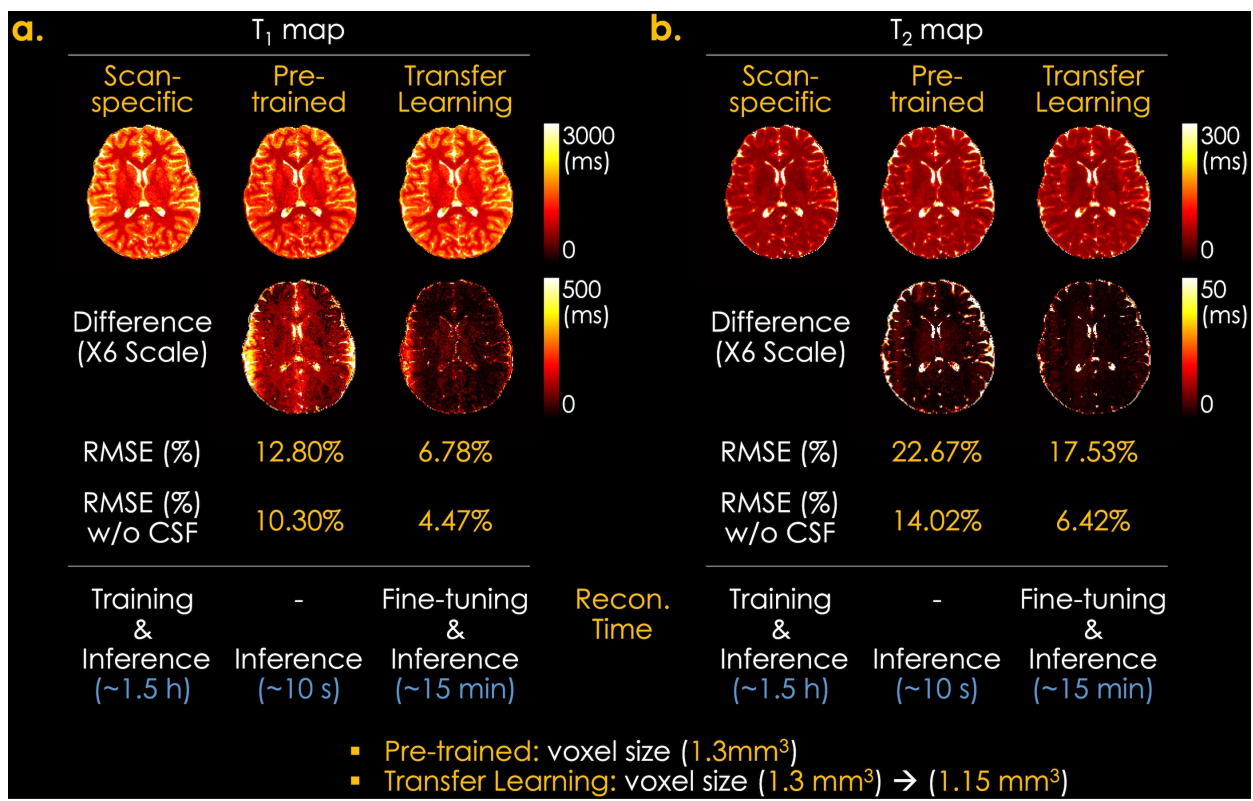

Supporting Information Figure S12. Reconstructed (a) $T_1$ and (b) $T_2$ maps using the proposed SSL-QALAS method with 3 different models (i.e., scan-specific, pre-trained, and transfer learning models). The pre-trained model was trained with the other subject's data, which has 1.3 mm$^3$ isotropic resolution, and the transfer learning model was initially trained with the other subject's data and fine-tuned with the target subject's data, which has 1.15 mm$^3$ isotropic resolution. The difference images show the difference between the reconstructed images of the scan-specific model and the ones of the pre-trained or transfer learning models. The reconstruction for each model takes about 1.5 h (scan-specific: training and inference), 10 s (pre-trained: inference only), and 15 min (transfer learning: fine-tuning and inference), respectively. Root mean square error (RMSE) was calculated between the scan-specific model and the pre-trained or transfer learning model.